\documentclass[conference]{IEEEtran}
\usepackage{graphicx,times,amsmath, amssymb}
\usepackage{fixltx2e} 
\usepackage{xspace}
\usepackage{url}
\usepackage{paralist}
\usepackage[caption=false,font=footnotesize]{subfig}
\usepackage{epstopdf}
\usepackage{graphicx}
\usepackage{balance}

\begin{document}

\title{
Structural Vulnerability Assessment of Electric Power Grids
}

\author{Yakup Ko\c{c}$^{1}$~~Martijn Warnier$^1$~~Robert E. Kooij$^{2,3}$~~Frances M.T. Brazier$^1$\\
$^1$Faculty of Technology, Policy and Management, Delft University of Technology, the Netherlands\\
$^2$Faculty of Electrical Engineering, Mathematics and Computer Science, Delft University of Technology, the Netherlands \\
$^3$TNO (Netherlands Organisation for Applied Scientific Research), the Netherlands

}
\date{}

\maketitle
\thispagestyle{empty}

\begin{abstract}

Cascading failures are the typical reasons of blackouts in power grids. The grid topology plays an important role in determining the dynamics of cascading failures in power grids. Measures for vulnerability analysis are crucial to assure a higher level of robustness of power grids. Metrics from Complex Networks are widely used to investigate the grid vulnerability. Yet, these purely topological metrics fail to capture the real behaviour of power grids. This paper proposes a metric, the effective graph resistance, as a vulnerability measure to determine the critical components in a power grid. Differently than the existing purely topological measures, the effective graph resistance accounts for the electrical properties of power grids such as power flow allocation according to Kirchoff laws. To demonstrate the applicability of the effective graph resistance, a quantitative vulnerability assessment of the IEEE 118 buses power system is performed. The simulation results verify the effectiveness of the effective graph resistance to identify the critical transmission lines in a power grid. 

\end{abstract}

\section{Introduction}
\label{sec_Introduction}

The electric power grid is crucial for economic prosperity, and national security. Disruptions to electrical power grids  paralyse the daily life in modern societies causing huge economical and social costs for these societies~\cite{Baldick2008}. The strong dependencies of other critical infrastructures such as telecommunications, transportation, and water supply on electric power grid amplifies the severity of large scale blackouts~\cite{Buldyrev2010, Eeten2011}. The key importance of the electric power grid to society encourages further research into sustaining power system reliability and developing new methods to evaluate and mitigate the risk of cascading blackouts.  



Power grid security is traditionally assessed relying on flow-based methods (e.g. $N-x$ contingency analysis~\cite{Baldick2009}). The flow-based methods model the continuous and discrete dynamics of the components and solve non-linear algebraic equations to determine the electric power flow distribution over the grid components. A complete security assessment based on flow-based methods requires evaluating a combinatory number of contingencies. However, this results in significant computational complexity and the associated computational time, enforcing power systems analysts to seek for alternatives~\cite{Baldick2009}. 

The strong connection between the topology and the robustness of a grid prompts a vulnerability assessment from a topological perspective. The recent advances in the field of complex networks~\cite{Barabasi99, Watts98} reveal its promising potential to investigate power grids vulnerability at the systems level from a topological perspective.

Topological investigation of electrical power grids demonstrates that power grids belong to the classes of small world~\cite{Watts98} networks, and they have scale free~\cite{Barabasi99} characteristics, suggesting that power grids have "hub" components having significant criticality compared to the rest of the network. These components are crucial for the grid~\cite{Latora2005}. Their removal weakens the system robustness significantly, or results in the largest possible damage in network performance.
Identifying these \emph{critical components} in a power grid is a major concern for power system security and attracts significant attention from the power system research community~\cite{Crucitti2005, Bompard2009, Dwivedi2009}. Identifying these critical components in a power grid in advance enables power grid operators to improve system robustness by monitoring and protecting these components continuously. 

To assess the structural vulnerability of power grids, several studies~\cite{Bompard2009, Bompard10, Arianos2009} have proposed extended topological approaches, while many others~\cite{Casals2007, Albert2004, Chassin2005} have used graph theoretical metrics including average shortest path length~\cite{Kim2007, Chen2006} and its derivatives such as efficiency~\cite{Crucitti2005}. These purely topological metrics consider only the structure of a power grid ignoring the electrical properties of the components in the network. However, in a power grid, the electric power is not governed only by the physical couplings (interconnections), but also by the electromagnetic couplings, and it flows through the network according to Kirchoff Laws. Therefore, the purely topological approaches considering only the topology fail to capture the real system behaviour. This paper proposes a metric, effective graph resistance, as a vulnerability measure to determine the critical lines in a power grid, while differently than the existing topological metrics, accounting for the power flow allocation according to Kirchoff Laws.

\section{Modelling Cascading Failures in Power Grids}
\label{sec_Modelling Cascading Failures in Power Grids}

A power grid is a multi-layered network that is composed of three functional parts: generation, transmission, and distribution. Power is provided from generation buses to distribution stations through the transmission buses that are all inter-connected via transmission lines. In a graph representation of a power grid, nodes represent generation, transmission, distribution buses, and substations, while links model the transmission lines and transformers. All the parallel transmission lines in the system are represented by an equivalent single link in a graph representation. Additionally, the links in a graph representation are weighted by the admittance (or impedance) value of the corresponding transmission line.

The electrical properties of a grid including impedances, voltage levels at each individual power station, voltage phase differences between power stations and loads at terminal stations control the power flow in the grid. This paper estimates the flow values for each component in the network by using linear DC power flow equations~\cite{Hertem2006} which is an approximation of nonlinear AC power flow equations~\cite{Grainger1994}. In a  DC model, the active power flow \emph{$f_{ij}$} through a transmission line \emph{$l_{ij}$} connecting node \emph{i} and node \emph{j} is related to the voltage phase values at both nodes \emph{i} and \emph{j} and the impedance of the line \emph{$l_{ij}$} as follows:
\begin{equation}\label{DC flow}
f_{ij}=\frac{\theta_{ij}}{x_{ij}}=b_{ij}\theta_{ij}
\end{equation}

\noindent where \emph{$\theta_{ij}$} is the voltage phase difference between node \emph{i} and node \emph{j}, \emph{$x_{ij}$} is the reactance, and \emph{$b_{ij}$} is the susceptance of \emph{$l_{ij}$}. The entire system can be modelled solely by:
\begin{equation}\label{DC flow summation}
P_{i}=\sum_{j=1}^{d} f_{ij}=\sum_{j=1}^{d} b_{ij} \theta_{ij}
\end{equation}

\noindent where \emph{$P_{i}$} is the real power flow at node \emph{i} and \emph{d} is the degree of node \emph{i}. In terms of matrices, Eq.~\eqref{DC flow summation} can be rewritten as:
\begin{equation}\label{DC flow equations}
\textbf{P}=\textbf{B}\boldsymbol\theta
\end{equation}

\noindent where \textbf{P} is the vector of real power injections, $\boldsymbol\theta$ contains the voltage phase values at each node, and \textbf{B} is the bus susceptance matrix in which \emph{$B_{ij}=-\frac{1}{x_{ij}}$} and \emph{$B_{ii}=\sum_{j=1}^{d}-B_{ij}$}. Since all the active power injections are known in advance, given the bus susceptance matrix \textbf{B}, the voltage phase values at each node can be calculated directly by using:
\begin{equation}
\label{DC flow equations_phases}
\boldsymbol\theta=\textbf{B}^{\textbf{-1}}\textbf{P}
\end{equation}

After obtaining the voltage angle values at each node, the power flow values through each line can be computed by using Eq.~\eqref{DC flow}.

The maximum capacity of a line is defined as the maximum power flow that can be afforded by the line. The flow limit of a transmission line is imposed by thermal, stability or voltage drop constraints~\cite{Glover2001}. This paper assumes that the maximum capacity $C_{l}$ of a line $l$ is proportional to its initial load $L_{l,in}$ by a tolerance parameter $\alpha_{l}$: $C_{l}=\alpha_{l} L_{l,in}$.

In a power grid, transmission lines are protected by relays and circuit breakers. A relay of a transmission line measures state variables (e.g. current), and compares them with a threshold value. When the threshold is violated, and this violation lasts long enough, the relay notifies a circuit breaker to trip the transmission line in order to prevent that the transmission line is permanently damaged due to e.g. overloading. This paper assumes a deterministic model for line tripping mechanism. A circuit breaker of a line $l$ trips at the moment the load $L_{l}$ of the line $l$ exceeds its maximum capacity $C_{l}$: $|L_{l}/C_{l}|>1$.

An initial outage of a component changes the balance of the power flow distribution over the grid and causes a redistribution of the power flow over the network. This dynamic response of the system to this triggering event might overload other parts in the network. The protection mechanism trips these newly overloaded components, and the power flow is again redistributed potentially resulting in new overloads. This cascade of failures continues until no more components are overloaded. 


\section{Complex Networks Preliminaries}
\label{sec_Preliminaries}

This section explains the relevant basic concepts from complex networks theory, introduces the effective graph resistance, and elaborates on how it is computed in electric power grids.

\subsection{Complex networks basics}
\label{subsec_preliminaries}
A network \emph{G(N,L)} consisting of a set of nodes \emph{N} and links \emph{L}, can fully be represented by its adjacency matrix. The \emph{adjacency matrix} $A$ of a simple, unweighted graph \emph{G(N,L)} is an $N \times N$ symmetric matrix reflecting the interconnection of the nodes in the graph. $a_{ij}=0$ indicates that there is no edge between nodes $i$ and $j$, otherwise $a_{ij}=1$. In case of a weighted graph, the network is represented by the weighted adjacency matrix $W$ where $w_{ij}$ corresponds to the weight of the line between nodes $i$ and $j$; a weight can be a distance, cost, or delay. 

The \emph{Laplacian matrix}~\cite{Mieghem2011} $Q$ is another way to fully characterize a graph, and defined as:
\begin{equation}\label{Laplacian}
Q=\Delta - A 
\end{equation}

\noindent where $\Delta$ is the diagonal matrix of strengths of $G$: $\delta_{i}$=$\sum_{j}^{N}{w_{ij}}$. Hence, the Laplacian can be constructed as: 
\begin{equation}
\label{QDesc}
  Q_{ij}=\begin{cases}
      
    \delta_{i}, 	& \text{if $i=j$}.\\
     -w_{ij},			& \text{if $i \neq j$ and $(i,j) \in L$}\\
     0,				& \text{otherwise}.

  \end{cases}
\end{equation}

\noindent where $\delta_{i}$ is the strength of node $\emph{i}$, and \emph{L} is the set of links in \emph{G}.

A \emph{walk} between pair of nodes \emph{i} and \emph{j} is a set of nodes and links that begins at node $i$ and ends at node $j$, while a \emph{path} $P_{ij}$ refers to a walk in which no nodes are revisited. The \emph{path length} $l(P_{ij})$ is the sum of the weights of constituent edges in the path $P_{ij}$. The shortest path length $l(P^{*}_{ij})$ is the minimizer of $l(P_{ij})$ over all $P_{ij}$. The \emph{average shortest path length} $l_{G}$ of a network $G$ is defined as:
\begin{equation}\label{CharPathLength}
l_{G}=\frac{1}{N(N-1)}\sum_{\substack{i \neq j \in G}}{l(P^{*}_{ij})}
\end{equation}

\subsection{Effective graph resistance in power grids}
\label{subsec_Computation of effective graph resistance}
Considering a network $G(L,N)$ with a Laplacian matrix constructed by the conductance values of the lines, the effective resistance~\cite{Mieghem2011} between a pair of nodes $i$ and $j$ $R_{ij}$ is the potential difference between these nodes when a unit current is injected at node $i$ and withdrawn at node $j$. The effective graph resistance is the sum of the individual effective resistances between each pair of nodes in the network. The effective graph resistance can be computed in two different ways: (a) by aggregating the effective resistances between each pair of nodes, and (b) by the eigenvalues of the Laplacian matrix of the grid. 


The required steps to compute the effective graph resistance based on pairwise effective resistances are (i) constructing the Laplacian matrix of the grid, (ii) determining the generalised inverse of the Laplacian matrix, (iii) computing effective resistances between each pair of nodes, and (iv) summing up the effective resistances. 

The Laplacian matrix of a power grid $Q$ reflects the interconnection of buses with transmission lines according to the description in Eq.~\ref{QDesc}. The weight $w_{ij}$ corresponds to the susceptance~\cite{Grainger1994} (i.e.\xspace inverse of reactance) value between nodes \emph{i} and \emph{j}. The Laplacian matrix constructed by the susceptance values is equivalent to the admittance matrix in the electrical power systems theory.

 The \emph{effective resistance} $R_{ij}$ between nodes \emph{i} and \emph{j} is computed as:
\begin{equation}\label{EffResistance}
R_{ij}=Q_{ii}^{+}-2Q_{ij}^{+}+Q_{jj}^{+}
\end{equation}

\noindent where $Q^{+}$ is the generalized inverse of $Q$ obtained by the Penrose pseudo-inverse operator~\cite{Moore1920}.

%


Subsequently, the \emph{effective graph resistance} $R_{G}$ of a power network is computed by summing up all the effective resistances between all pairs in a network. 
\begin{equation}\label{EffGraphResistance}
R_{G}={\sum_{i=1}^{N}}{\sum_{j=i+1}^{N} {R_{ij}}}
\end{equation}


Another way to compute the effective graph resistance of a power grid requires computation of the eigenvalues of the Laplacian matrix of the grid. This approach requires summing the reciprocal of the eigenvalues:
\begin{equation}\label{EffGraphResistanceEigenValues}
R_{G}=N{\sum_{i=1}^{N-1}} \frac{1}{\mu_{i}}
\end{equation}

 \noindent where $\mu_{i}$ is the $i^{th}$ eigenvalue of the Laplacian matrix. This methodology is computationally more efficient, but it does not give any insight into the individual effective resistances between pairs of buses.
 
\begin{figure}[htb]
\centering
\includegraphics[scale=0.5]{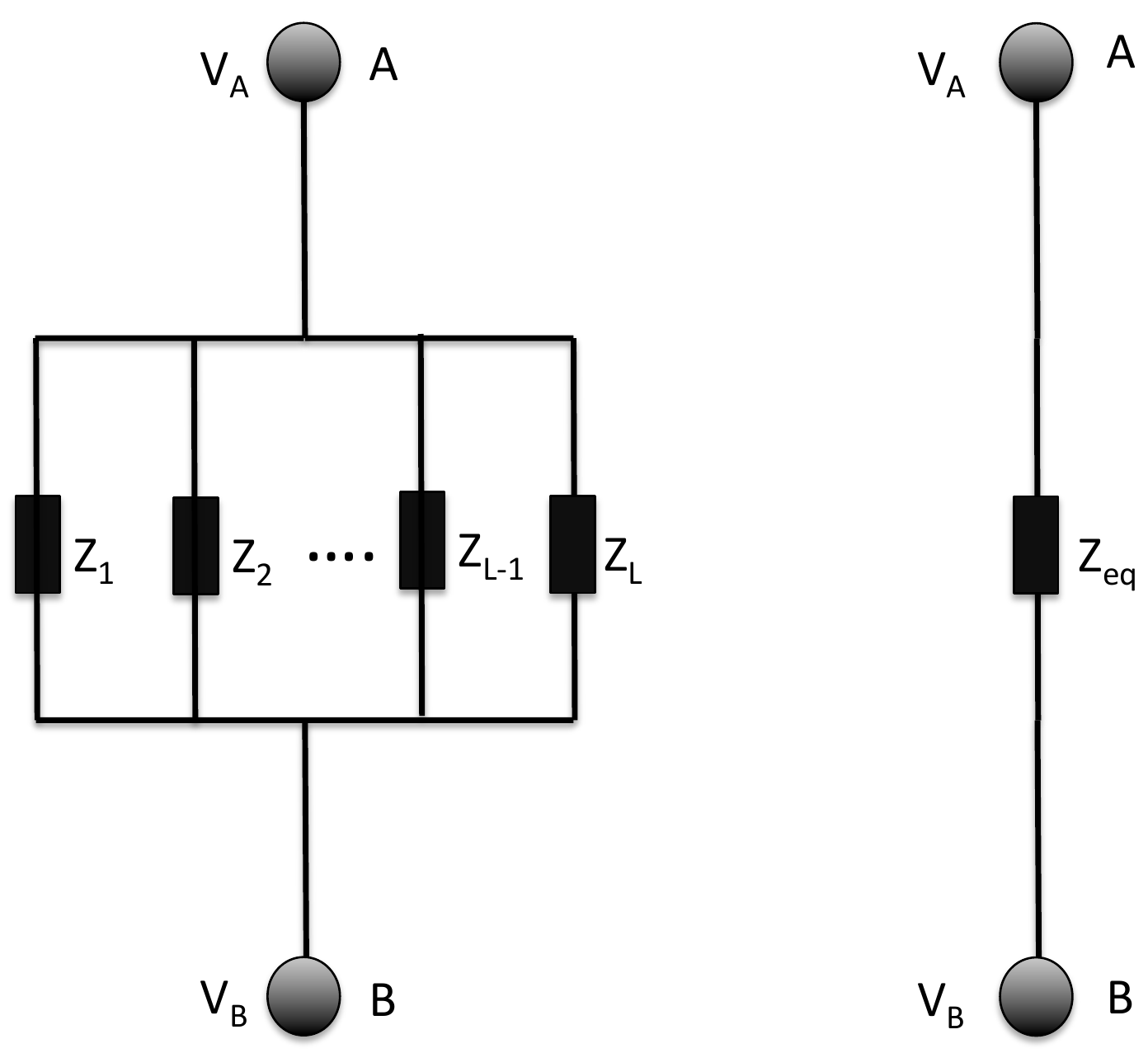}
\caption{Multiple $L$ branches between nodes $A$ and $B$ with impedances $Z_{1},Z_{2},..,Z_{L}$ can be replaced by one single branch with an impedance value $Z_{eq,AB}$. $Z_{eq,AB}$ is related to the potential difference between $A$ and $B$ as: $V_{A}-V_{B}=IZ_{eq,AB}$. Assuming a unit electric current $I$, the equivalent impedance equals the potential difference between nodes $A$ and $B$, and also, per definition, the effective resistance between nodes $A$ and $B$.}
\label{fig:EqImpedance}
\end{figure}

 In a DC model of a electrical power grid, the effective resistance $R_{ij}$ between buses $i$ and $j$ is equal to the equivalent impedance $Z_{eq,ij}$ between these buses. Fig.~\ref{fig:EqImpedance} illustrates the case.
 
\section{Effective Graph Resistance as a Metric to Assess Grid Vulnerability}
\label{sec_EGRRobMetric}

In purely topological approaches such as shortest path length, electrical power is assumed to follow the shortest or the most efficient path. Relying on this assumption, a shortest path between a pair of nodes is determined. The criticality of a transmission line is then quantified based on the impact of its removal on the network average shortest path length. However, electric power is not governed only by topological, but also by the electrical properties of the grid. The electric power flows through all the possible paths rather than following one single path (e.g. the shortest path). Consequently, these purely topological approaches, that consider only the topology of a grid and ignore the electric characteristics, fail to capture the real behaviour of the grid in terms of robustness and vulnerability. 

In power grids, utilization of multiple paths precludes the existence of a physical shortest path between two buses. However, the concept of \emph{effective resistance} makes it possible to determine a distinct \emph{electrical path} between two nodes by conceptually replacing the multiple paths between two nodes with a single equivalent path. The effective resistance between two nodes is the total cost incurred to transfer electric power between these nodes. Consequently, the effective resistance is the (real) electrical path length between two nodes, and can replace its purely topological counterpart (the shortest path length) for a realistic vulnerability analysis of power grids. 
%

\begin{figure*}
\begin{center}	
	\includegraphics[width=.35\textwidth]{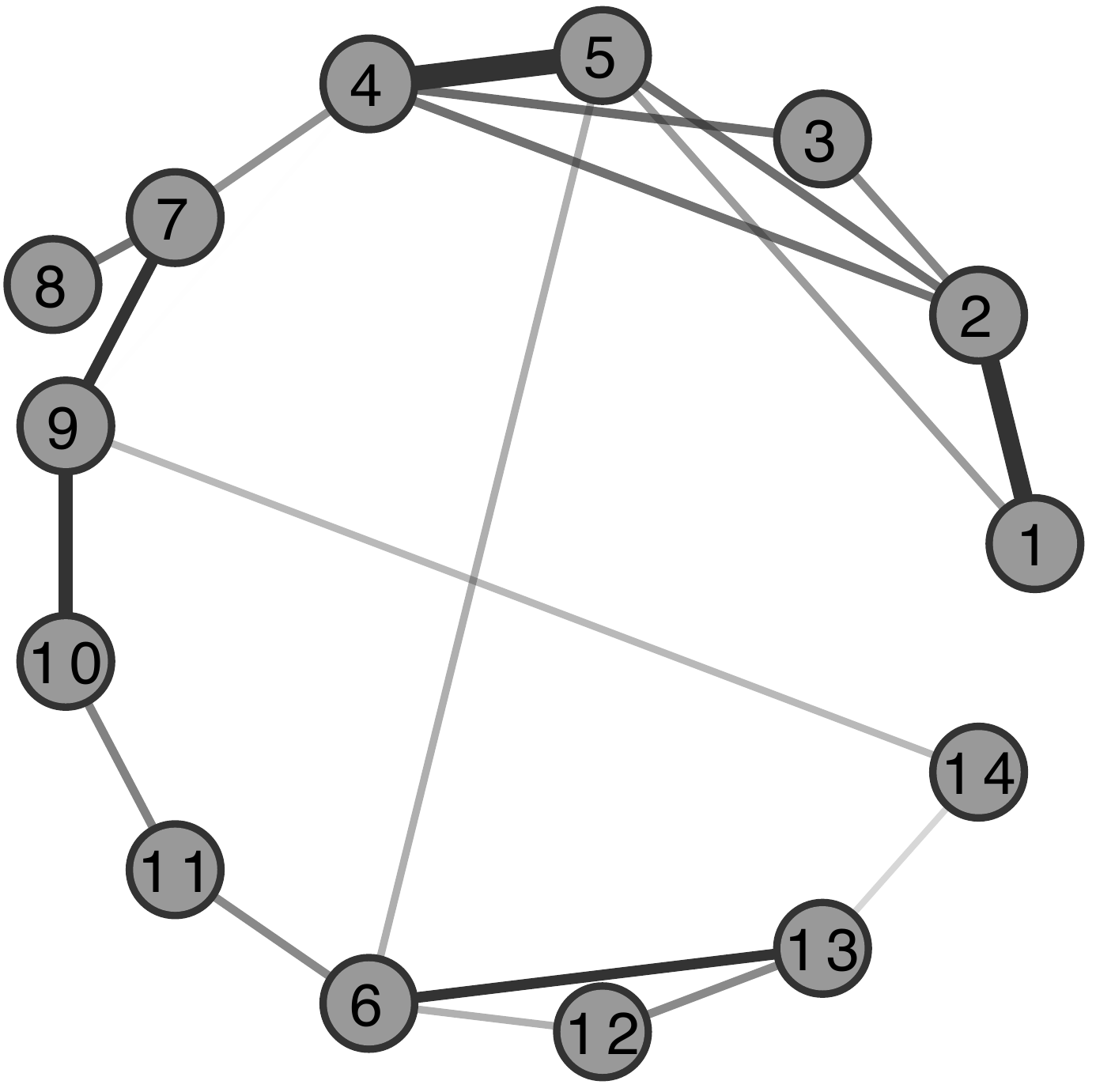}
	\label{fig:IEEE30Physical}
	\includegraphics[width=.35\textwidth]{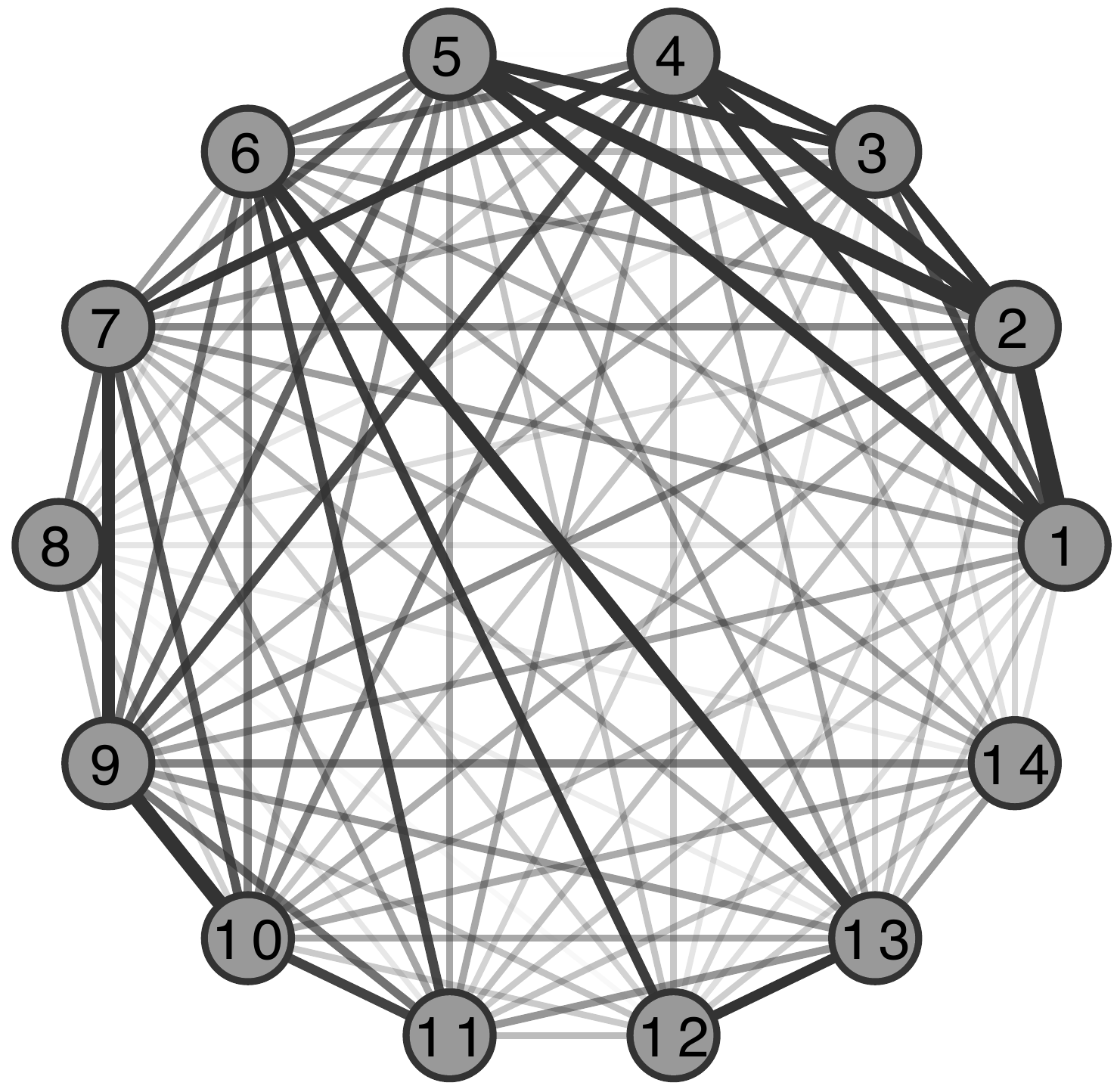}
	\label{fig:IEEE30Electrical}
\caption{The physical (left) and the electrical (right) topology of the IEEE 14 power system. In the physical topology the conductances, and in the electrical topology the effective conductances (i.e.\xspace $1/R{ij}$), are used as weights for a better illustration. In the physical topology, a relatively thicker and more visible line corresponds to a stronger connection (i.e.\xspace a smaller effective resistance), while a relatively thinner and less visible line corresponds to a relatively weaker connection (i.e.\xspace larger effective resistance).}
\label{fig:IEEE30}
\end{center}
\end{figure*}

The concept of effective resistance makes it possible to construct the \emph{electrical topology} of a power grid. An electrical topology of a power grid shows the electrical connections between buses, rather than the physical connections as a physical topology does. In an electrical topology of a power grid, the nodes represent generation, transmission, distribution buses, and substations while a link between nodes ${i}$ and ${j}$ corresponds to the effective resistance $R_{ij}$. Fig.~\ref{fig:IEEE30} shows the physical and the electrical topology of the IEEE 14 power system~\cite{TestCaseRef}. 

%
In a physical power grid topology, the existence of parallel paths between two nodes, and a homogeneous distribution of their impedance values result in a smaller effective resistance between these two nodes. The number of parallel paths in the physical topology refers to the number of redundant (backup) paths. In case of a failure in one of the paths between two nodes, the power flow carried by the rendered path is distributed over the backup paths. Therefore, a higher number of backup paths implies a more robust network against cascading failures due to line overloads. On the other hand, a relatively more homogeneous distribution of the impedance values results in a relatively more homogeneous distribution of power flow over these parallel paths increasing the robustness of the power grid against cascading failures~\cite{Koc2013, Koc2013_2}. Therefore, a power grid with a relatively smaller effective graph resistance implies a relatively more robust power grid with respect to cascading failures. Ko\c{c} et al.~\cite{Koc2013_4} verifies the relationship between $R_{G}$ and the robustness of a power grid.


From an electrical topology point of view, a relatively small $R_{G}$ results in relatively strong local electrical connections (e.g. electrical connection between buses 1 and 2 in the electrical topology in Fig.~\ref{fig:IEEE30}) and relatively strong remote electrical connections (e.g. electrical connection between buses 6 and 13 in the electrical topology in Fig.~\ref{fig:IEEE30}) between buses. The strong local electrical connections in a part of a power grid allows a better accommodation of power flow in that area, increasing the local robustness~\cite{Koc2013, Koc2013_2}. On the other hand, the strong remote electrical connections between distant buses enable transfer of electrical power flow from a region of the grid to another region resulting in a better accommodation of power flow over the entire power grid. In case of a congestion in one part of the grid (e.g. as a result of a failure), the excess power in the associated part is easily transferred to the rest of the grid, enabling the utilization of the redundant capacity in the rest of the grid. Accordingly, relatively stronger electrical connections between buses (i.e. a relatively smaller effective graph resistance) allow a better accommodation of power flow and increase the ability of a power grid to distribute the excess power over the rest of the network ensuring a higher tolerance against local failures.


 The effective graph resistance locates the transmission lines that are contributing most to the electrical connections in a power grid. Removal of these lines reduces the ability of a power grid to better accommodate power flow, hence, the attack tolerance of the grid. Accordingly, these lines are the lines with the highest criticality in a power grid and a robust operation of a power grid requires continuous monitoring and protection of these lines.

\section{Use Case:Identifying the critical lines in a power grid}
\label{sec_Use Case}

This section demonstrates how the effective graph resistance is used as a metric to asses power grid vulnerability, and to determine the critical components for the system.

For a quantitative criticality analysis, the IEEE 118 buses power system~\cite{TestCaseRef} is considered. To assess the criticality of a transmission line based on $R_{G}$, this section deploys an analogous approach to the one given in~\cite{Latora2005, Mieghem2011_2}: the criticality of a transmission line $l$ in a grid $G$ is determined by the relative increase in the effective graph resistance $\Delta R_{G}^{l}$ that is caused by the deactivation of line $l$:

\begin{equation}\label{Criticality}
\Delta R_{G}^{l}=\frac {R_{G-l}-R_{G}}{R_{G}} 
\end{equation}

\noindent where $R_{G}(G-l)$ is the effective graph resistance of the grid that is obtained from $G$ by removing $l$. 

\begin{figure*}
\centering
\includegraphics[scale=0.5]{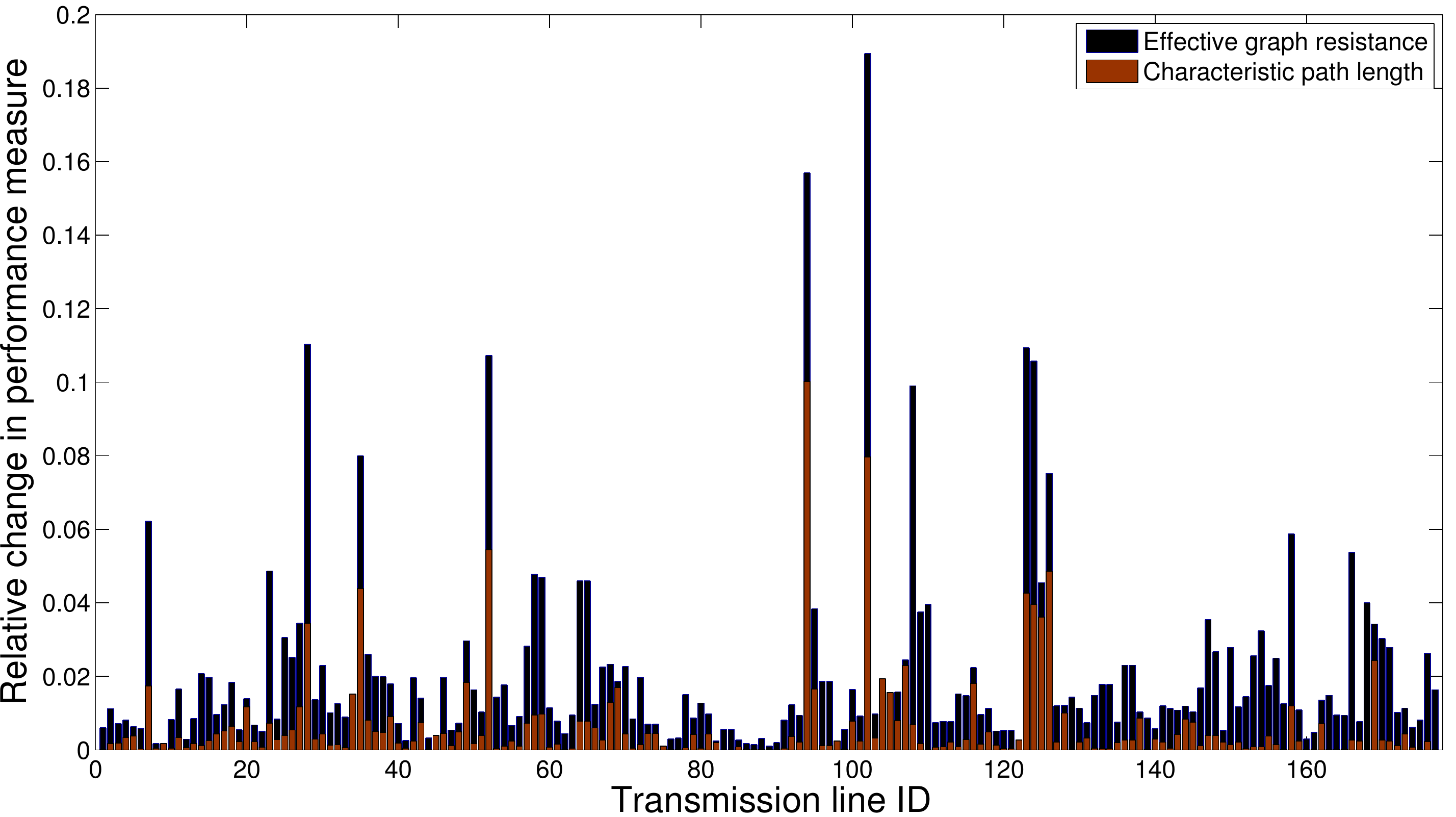}
\caption{Relative increase in effective graph resistance and in average shortest path for IEEE 118 test case}
\label{fig:IEEE118LineAnalysis}
\end{figure*}

The original grid $R_{G}$, and the impact of each individual line removal on $R_{G}$ are computed. By substituting these values in Eq.~\ref{Criticality}, the impact of each transmission line on the grid robustness ($\Delta R_{G}^{l}$) is assessed. At the same time, the criticality of each transmission line is also quantified based on $l_{G}$ ($\Delta l_{G}^{l}$) by following the same approach. Fig.~\ref{fig:IEEE118LineAnalysis} shows the results for all transmission lines in the IEEE 118 power system, while Table~\ref{tab:10 critical lines} shows the 10 most critical lines identified by $R_{G}$ and $l_{G}$.

\begin{table}
\centering
\caption{Most critical 10 transmission lines in IEEE 118 power system based on $R_{G}$ and $l_{G}$}
\label{tab:10 critical lines}
\begin{tabular}{c c c c} \hline
 	Line ID 	 & $\Delta R_{G}^{l}$(\%) & Line ID & $\Delta l_{G}^{l}$(\%)\\
\hline
$l_{65-68}$			  	& 18.94		& $l_{38-65}$		& 10.02\\
\hline
$l_{38-65}$			  	& 15.70		& $l_{65-68}$		& 7.97\\
\hline
$l_{23-24}$			  	& 11.03		& $l_{30-38}$		& 5.44\\
\hline
$l_{68-81}$ 				& 10.94		& $l_{82-83}$		& 4.86\\
\hline
$l_{30-38}$ 				&10.70			& $l_{8-30}$		& 4.39\\
\hline
$l_{80-81}$ 				& 10.57		& $l_{68-81}$		& 4.27\\
\hline
$l_{70-71}$			  	& 9.90			& $l_{80-81}$		& 3.95\\
\hline
$l_{8-30}$ 				& 8.30			& $l_{77-82}$		& 3.61\\
\hline
$l_{82-83}$ 				&7.52			& $l_{23-24}$		& 3.44\\
\hline
$l_{8-5}$ 				   & 6.22			& $l_{103-110}$	& 2.43\\
\hline
\end{tabular}
\end{table}

In Fig.~\ref{fig:IEEE118LineAnalysis} and Table~\ref{tab:10 critical lines}, the line criticality analysis based on $R_{G}$ shows that the transmission line with the ID of 104 (connecting bus 65 to bus 68) is the most critical line for the IEEE 118 power system. Deactivation of line 104 causes nearly a 19\% increase of the effective graph resistance of the grid. On the other hand, the analysis based on the average shortest path length suggests that the transmission line with the ID of 96 (connecting bus 38 to bus 65) is the most critical line and its removal results in nearly 10\% increase of average shortest path length of the grid topology.

The top 10 most critical lines according to $R_{G}$ are almost the same as the top 10 lines based on $l_{G}$. However, these lines have different rankings in their criticality. The purely topological $l_{G}$ ignores the electrical properties and identifies the criticality of transmission lines purely based on the centrality of their location in the grid. On the other hand, $R_{G}$ incorporates the impact of the electrical properties of transmission lines in addition to the importance of their central locations. This makes the difference between the two approaches and results in different ranking of importances of these components.

To validate the results from Table~\ref{tab:10 critical lines}, the IEEE 118 power system is attacked by removing the critical lines identified by $R_{G}$ and $l_{G}$, and the damage is quantified. The simulations are performed by MATCASC~\cite{Koc2013_3}, a MATLAB based cascading failures analysis tool implementing the model in Sec.~\ref{sec_Modelling Cascading Failures in Power Grids}. The damage caused by the cascade is quantified in terms of normalized served power demand: served power demand divided by the total power demand in the grid. The attack vulnerability of the grid differs under various operative states for the same attack. To capture the vulnerability of the system under different operative states, the normalized served power demand is averaged over 100 random instances of power demand, varying in the interval of [$P_{d}$, 3$P_{d}$] ($P_{d}$ is the base power demand given in~\cite{TestCaseRef}).

\begin{figure}[htb]
\centering
\includegraphics[scale=0.30]{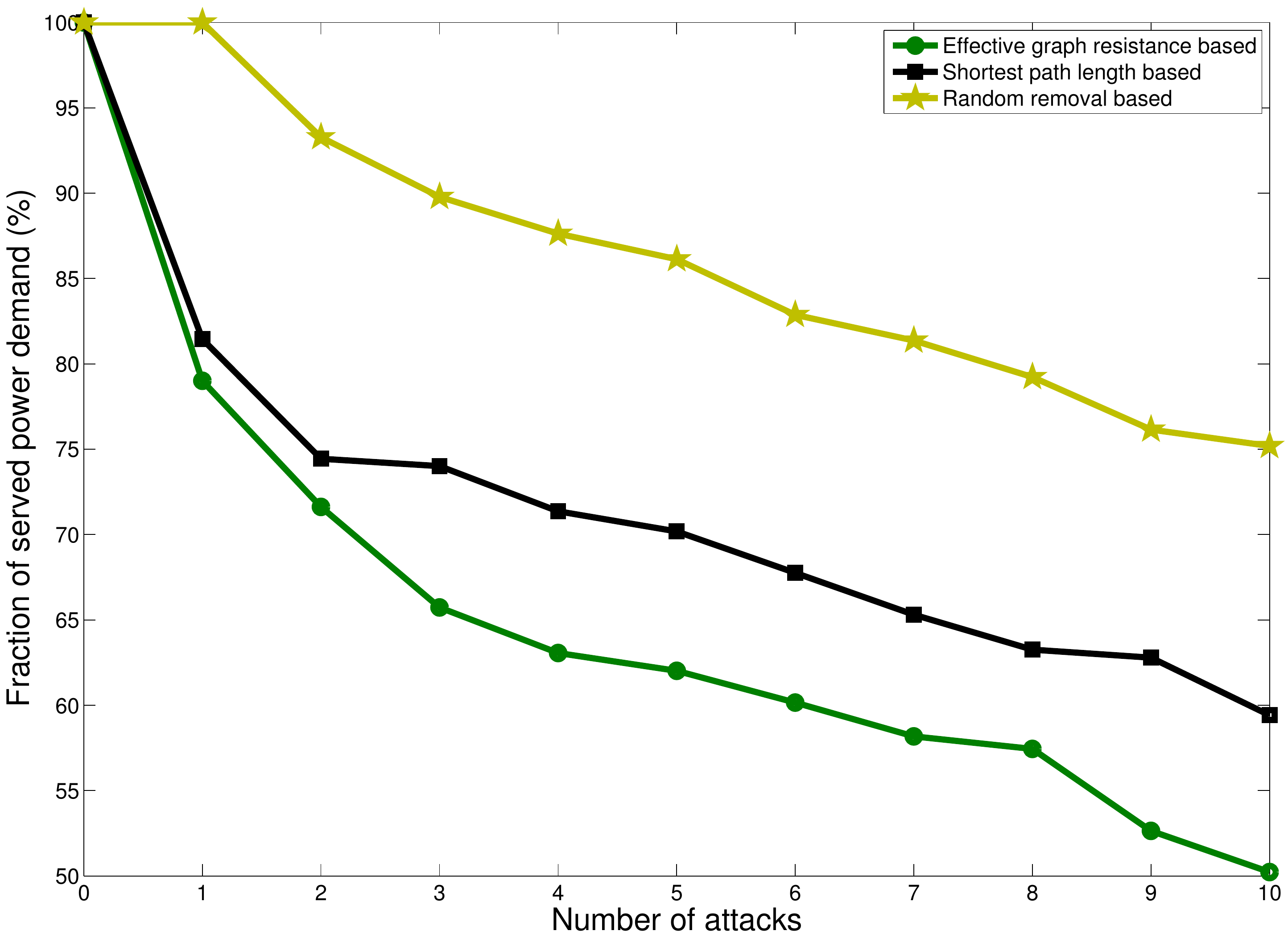}
\caption{Effectiveness of attacks based on effective graph resistance, average shortest path length, and random removals for IEEE 118 power system}
\label{fig:IEEE118AttackSeqAnalysis}
\end{figure}

Initially, the top 10 critical lines identified by $R_{G}$ (see Table~\ref{tab:10 critical lines}) are attacked one after the other. After each successive attack, the served power demand in the grid is quantified. The same analysis is performed by attacking the top 10 critical lines based on $l_{G}$ (see Table~\ref{tab:10 critical lines}), and 10 lines based on random removals. Fig.~\ref{fig:IEEE118AttackSeqAnalysis} shows how the fraction of served power demand decreases after each successive attack. The largest damage results after an attack strategy based on $R_{G}$ validating the effectiveness of $R_{G}$ to identify the critical transmission lines in a power grid. 

\section{Conclusion and discussion}
\label{sec_Conclusion}
This paper proposes the \emph{effective graph resistance} $R_{G}$ as a global vulnerability measure for power grids. Based on $R_{G}$, the critical transmission lines in a power grid are determined. The proposed metric $R_{G}$ serves as a better tool to assess the grid vulnerability compared to traditional average shortest path length by incorporating the electrical properties of power grids such as power flow allocation according to Kirchoff Laws. The proposed approach is applied on the IEEE 118 power system to determine the critical components. Results are compared to the traditional topological metric average shortest path length. Simulation results verify the effectiveness of $R_{G}$ as a measure to assess power grid vulnerability.  

\balance
 \subsection*{{\bf Acknowledgements}}
 This research is partially funded by the NWO project \emph{RobuSmart}, grant number 647.000.001, and the KIC EIT projects SES 11814 \emph{European Virtual Smart Grid Labs}.

\bibliographystyle{IEEEtran}


\end{document}